\documentclass[12pt,a4paper]{article}
\pdfoutput=1
\usepackage{jheppub}
\usepackage{graphicx}
\usepackage{lipsum}
\usepackage{youngtab}
\usepackage{tikz}
\usepackage{tikz-cd}
\usepackage{braids}
\usepackage{amsmath,amsfonts}
\usepackage{slashed}
\usepackage{enumitem}
\usepackage[all]{hypcap}
\allowdisplaybreaks

\usepackage{amsthm}

\theoremstyle{definition}

\def\Z{\mathbb{Z}}
\def\mod{\text{mod~}}
\graphicspath{ {./Figs/} }
\usepackage{xspace}

\newcommand{\be}{\begin{equation}}
\newcommand{\ene}{\end{equation}}
\newcommand{\T}{T}
\newcommand{\Tp}{{T'}}
\newcommand{\avg}[1]{\left\langle#1\right\rangle}

\usepackage{amscd,amsxtra,mathrsfs,graphics,graphicx,amsthm,epsfig,bm,longtable,float,color,tikz,mathtools,xfrac,footnote,rotating,lscape,wrapfig}
\usepackage{simpler-wick}
\usepackage[makeroom]{cancel}
\usepackage[utf8]{inputenc}
\usepackage[T1]{fontenc}
\usepackage{bm}
\usepackage{physics}
\usepackage{caption,subcaption}
\usetikzlibrary{arrows}
\usepackage{empheq}
\usepackage{array}   
\newcolumntype{C}{>{$}c<{$}} 

\def\SO{\text{SO}}

\def\SU{\text{SU}}

\def\su{\mathfrak{su}}

\def\U{{\text{U}(1)}}

\def\Pin{\text{Pin}}

\def\GCD{\text{gcd}}
\def\LCM{\text{lcm}}

\def\UU{\text{U}}
\def\u{\mathfrak{u}}

\def\Tr{\text{Tr}}
\def\diag{\text{~diag}}

\def\ee{\text{e}}

\newcommand{\QCD}{\text{QCD}}

\title{Comments on QCD$_3$ and anomalies with fundamental and adjoint matter}

\date{}
\author[a]{Nakarin Lohitsiri}
\author[a]{and Tin Sulejmanpasic}

\affiliation[a]{Department of Mathematical Sciences, Durham University}
\emailAdd{nakarin.lohitsiri@durham.ac.uk}
\emailAdd{tin.sulejmanpasic@durham.ac.uk}
\abstract{'t Hooft anomaly matching is powerful for constraining the low energy phases of gauge theories. In 3d one common anomaly is the parity anomaly in a $\T$-symmetric theory where one cannot gauge the global symmetry group without breaking the time-reversal symmetry. We find that a $\T$-symmetric $\SU(N)$ gauge theory with either fermionic or bosonic matter in the fundamental representation of the gauge group has a parity anomaly between the flavor group and $\T$-symmetry provided that there is also a massless Majorana fermion in the adjoint representation of the gauge group. In particular, there is always a mixed anomaly between $\T$ and $\U$ baryon symmetry. We then analyze the parity anomaly in this theory, together with the more recent mod 16 time-reversal anomaly, and give some free fermion proposals as candidates for the low energy phases consistent with the anomalies. We make brief comments about the large $N$ limit and the $\T$-broken regimes in the conclusion as well as related anomalies in 4d. }
\begin{document}
\maketitle

\section{Introduction and Outline}
\label{sec:intro}

There has been a resurgence of interest in 3d gauge theories in recent years, especially in those without supersymmetry  \cite{Karch:2016sxi,Seiberg:2016gmd,Radicevic:2016wqn,Karch:2016aux,Hsin:2016blu,Benini:2017dus,Cordova:2017kue,Komargodski:2017keh,Gomis:2017ixy,Karch:2018mer,Aitken:2019shs}. The activity was in part fueled by the discovery of new 't Hooft anomalies which involve finite symmetry groups, generalized global symmetries and their connection to symmetry protected topological phases  (see e.g. \cite{Gaiotto:2014kfa,Gaiotto:2017yup,Gaiotto:2017tne,Komargodski:2017dmc} and references therein). Along a similar line of enquiries, in this work we discuss 3d $\SU(N)$ gauge theories with a single massless Majorana fermion in the adjoint representation of the gauge group and matter fields (either bosonic or fermionic) in the fundamental representation which can generically be massive (we will henceforth call this theory QCD$_3(\textbf{adj}/\textbf{f})$). Further we are interested in time-reversal-symmetry-preserving theories. The fundamental matter has flavor symmetries, which includes the $\U$-baryon symmetry. Without the massless Majorana adjoint fermion the theory has no 't Hooft anomalies, which is obvious by the fact that one can deform the theory to a trivially gapped theory by condensing scalar fundamentals, and Higgsing the gauge group completely. The situation is different in the presence of the Majorana-adjoint fermion, as the $\T$-symmetry prohibits it from being gapped out. The Higgs regime now has gapless fermions, and there is no obvious way to gap out the theory trivially.

In fact, as we will show, the presence of a Majorana-adjoint causes the system to have a mixed flavor--time-reversal symmetry anomaly. This is even true for one flavor of fundamental matter, either bosonic or fermionic, so that the flavor symmetry is only a $\U$-baryon symmetry. In this case the mixed $\T$--$\U$ anomaly is the famous parity anomaly\footnote{The parity anomaly is usually viewed as the absence of $\T$-symmetry in the presence of dynamical gauge fields at the quantum level. However here we take the point of view of the parity anomaly as an 't Hooft anomaly.} \cite{Niemi:1983rq,Redlich:1983dv,Redlich:1983kn}, which we will review in Sec.~\ref{sec:parity-anomaly}. The presence of gapless fermions in the Higgs regime can be restated as matching the parity anomaly in the IR in the simplest possible fashion, namely, with free Dirac fermions with non-zero baryon charge.  Interestingly, the anomaly is still present even when the fundamental matter is heavy\footnote{In this case the theory reduces to Super Yang-Mills in 3d which which was argued to have a TQFT \cite{Gomis:2017ixy} along with the massless goldstino \cite{Witten:1999ds}.}.

When we increase the number of flavors to $N_f>1$, the faithful global symmetry group which acts on gauge-invariant operators becomes $\UU(N_f)/\Z_N$. There is still a mixed flavor--$\T$ anomaly, but more care must be taken to characterize it. This characterization will be detailed in Sec. \ref{sec:UNf-parity} by defining two mod 2 indices as generalizations of Chern--Simons level mod 1 for the classic parity anomaly. Moreover, the mixed anomaly between $\T$ and $\U$ baryon symmetry persists. This is different from $\SU(N)_0$ gauge theory with only even $N_f$ fundamental fermions where there is a mixed $\UU(N_f)/\Z_N$--$\T$ anomaly but no mixed anomaly between $\T$ and the $\U$ baryon subgroup. Such a theory is studied in the large $N$ limit in \cite{Armoni:2019lgb} where the non-abelian part of the flavor symmetry group could be spontaneously broken, consistent with the anomalies. More interestingly, there can also be 't Hooft anomaly in the $\T$ symmetry itself, which is determined by a mod 16 index. We will see that, in QCD$_3(\textbf{adj}/\textbf{f})$, there are two such $\T$ symmetries which differ by charge conjugation, and therefore two mod 16 indices can be defined, as we explain in Sec. \ref{sec:mod-16}, giving further constraints on the possible IR phases.  The simplest way to match all these anomalies in the IR is with free fermions charged under the global symmetry, together with some additional neutral fermions. The neutral fermions are needed to satisfy one of the $\mod$ 16 anomalies. One can even see them arising explicitly in the completely Higgsed regime for the theory with scalar fundamentals. We will discuss this particular example in some details in Sec. \ref{sec:QCDf-adj} to illustrate this phenomenon in particular as well as anomaly matching in general.

The rest of the paper is organized as follows. In the rest of Sec.~\ref{sec:QCDf-adj}, we consider  QCD$_3(\textbf{adj}/\textbf{f})$ with a general number of fundamental fields. There, we discuss its global symmetry structure and analyze the anomalies in details for both the fermionic fundamental and the scalar fundamental cases. Then, motivated by the totally Higgs phase of the scalar theory, we also take the free fermion phase in the IR as a viable option and propose some free fermion configurations consistent with both the parity anomaly and the mod 16 anomalies. We summarize our investigation and discuss several alternatives for the IR phase, as well as indicating some interesting directions for further study, in Sec. \ref{sec:conclusion}. Some lengthy computations whose results are used in the main text are collected in the Appendices.

\section{The parity anomaly}
\label{sec:parity-anomaly}

In this section we will review the parity anomaly, which can be thought of as the mixed anomaly between a $\U$ global symmetry and the time reversal symmetry $\T$ in 3d. The anomaly is epitomized in the single Dirac fermion in 3d, given by the Lagrangian
\be
\mathcal L=\bar\psi i\slashed \partial \psi\;,
\ene
where $\psi$ is a Dirac fermion in 3d, and $\slashed \partial=\gamma^\mu\partial_\mu$, where $\gamma^\mu$ are the 3d gamma-matrices which we take in Minkowski space to be
\begin{equation}\label{eq:gamma_matrices}
\gamma^0= i\sigma^2\;,\quad \gamma^1=\sigma^1\;, \quad \gamma^2=\sigma^3\;.
\end{equation}
Note that the gamma-matrices are all real. We define the time-reversal symmetry $\T$ to be\footnote{Note that in high-energy physics the time reversal symmetry is usually taken not to complex conjugate the spinor $\psi$. The definition of $\T$ here corresponds in the high-energy community to $CT$. We find the current definition more convenient as the basic $\T$-symmetry, as the complex conjugation comes from the anti-unitarity of $T$, and we prefer not to undo the complex conjugation by charge conjugation symmetry.} 
\be\label{eq:T_sym_dirac}
\T:\; \psi(x^0,x^1,x^2)\rightarrow \gamma^0\psi^*(-x^0,x^1,x^2)\;.
\ene
Now the mass term in the Lagrangian is given by 
\be
\mathcal L_m=im\bar\psi\psi\;.
\ene
The imaginary ``$i$'' is important to make the Minkowski-space action real. However it is easy to show that because of this ``$i$'' the action of an anti-unitary $\T$-symmetry gives
\be
\T: \mathcal L_m \rightarrow - \mathcal L_m\;,
\ene
and hence the mass term breaks time-reversal symmetry. 

If we now put background gauge field for the $\U$ symmetry $\psi\rightarrow e^{i\alpha}\psi$, we need to promote
\be
\partial_\mu\rightarrow D_\mu=\partial_\mu - i\mathcal A_\mu\;.
\ene
It is then well known that the Dirac determinant cannot be regulated without breaking the $\T$-symmetry. The partition function is given by
\be
Z={\det}_{reg} \slashed D\;,
\ene
where $reg$ refers to the regulated determinant.
Under the $\T$ symmetry the single fermion partition function transforms as
\be
\T:\;Z\rightarrow Z e^{\pm\frac{i}{4\pi}\int_X \mathcal Ad\mathcal A}\;,
\ene
where the sign in the exponent depends on the sign of the regulator mass and where $X$ is the 3d  space-time manifold. We will always pick the lower sign, so that the action transforms as
\be
\Delta S= \frac{1}{4\pi}\int_X \mathcal Ad\mathcal A\;.
\ene
The important point is that the non-invariance under $\T$ cannot be fixed with a local counterterm, but it can be fixed by postulating an auxiliary bulk $\Sigma$ whose boundary is the space-time manifold $X=\partial \Sigma$ equipped with an invertible quantum field theory with the action
\be
S_{SPT}=\frac{1}{8\pi} \int_{\Sigma} \mathcal F\wedge \mathcal F\;,
\ene
where SPT stands for Symmetry-Protected Topolological phase -- another name for the gapped bulk theory. This is an an example of anomaly inflow where an 't Hooft anomaly is captured by an SPT phase in one dimension higher. 

The inability to preserve the $\T$-symmetry  in the presence of the $\U$ gauge field is a reflection of a mixed 't Hooft anomaly between the $\T$ symmetry and the $\U$ symmetry. This means that it is impossible to trivially gap the theory without breaking one of these two symmetries explicitly, for example by adding a bare Dirac mass term which breaks the $\T$ symmetry.

We will now discuss a parity anomaly involving a symmetry group $\UU(N_f)/\Z_N$, as we will need it for the discussion of the $\QCD_3(\textbf{adj}/\textbf{f})$ theories, the subject of study for this paper.

\subsection{The $\UU(N_{f})/\Z_{N}$ parity anomaly}
\label{sec:UNf-parity}

First consider a Dirac fermion in a representation $\mathbf R$ of simple Lie gauge group $G$. Let $A$ be the gauge field in the fundamental representation of $G$ coupled to such a fermion. The claim is that the measure now transforms as
\be
\Delta S = \frac{2d(\mathbf{R})}{4\pi} \int \tr\left(AdA-\frac{2i}{3}A^3\right)
\ene
where $d(\mathbf{R})$ is the Dynkin index\footnote{The appearance of the Dynkin index can be traced to the vacuum polarization diagram with a fermion loop inside. Such a diagram will feature the generators the trace $\tr_{\mathbf R}(T^aT^b)=d(\mathbf R) \delta^{ab}$, where $T^a$ are generators of $G$.} of the representation $\mathbf{R}$ of $G$. We normalize the Dynkin index such that $d(\textbf{f})=1/2$ where $\textbf{f}$ stands for the fundamental representation of $G$. This is equivalent to informally stating that the measure ``induces'' a Chern--Simons term with level $-d(\textbf{R})$ which we will adopt for the rest of the paper.\footnote{In the case of a Majorana fermion instead of a Dirac fermion, the induced Chern--Simons level is $-d(\textbf{R})/2$.} 

Now we want to consider a free massless fermion in a global symmetry group \footnote{The full symmetry group of $\text{dim}(\textbf{R})$ free massless Dirac fermions is certainly larger than this. We only focus our attention to its subgroup $\UU(N_f)/\Z_N$ for the purpose of anomaly matching where the global symmetry of theories in the UV is $\UU(N_f)/\Z_N$.}
\begin{equation}
  \label{eq:global-symmetry-free-fermion}
  F = \frac{\UU(N_f)}{\Z_N} \cong \frac{\SU(N_f) \times \U}{\Z_N \times \Z_{N_f}}.
\end{equation}
To classify representations of such a symmetry group, consider the fundamental representation of $\UU(N_f)$ labeled by $\psi^{I}$ where $I=1,\dots, N_f$. To build an arbitrary representation of $\UU(N_f)$ we can take direct products of fundamental representations and symmetrize or anti-symmetrize them with respect to the fundamental indices as usual. The difference between $\UU(N_f)$ and $\SU(N_f)$ is reflected in the fact that when $N_f$ indices are anti-symmetrized, they leave behind $N_f$ units of a $\U$ charge\footnote{We normalize the $\U$ charge such that the fundamental representation of $\UU(N_f)$ has a unit $\U$ charge. }. So for example completely anti-symmetric $N_f$-box representation has a $\U$ charge $N_f$. In other words to fully specify the representation of $\UU(N_f)$ we must specify a representation of $\SU(N_f)$ and the $\U$ charge. Let $\textbf R_q$ be a representation of $\UU(N_f)$ where $\textbf R$ is a representation of $\SU(N_f)$ and $q$ specifies the $\U$ charge. Note that we must have that 
\be
q-|\textbf R|=0\bmod N_f\;,
\ene 
where $|\textbf R|$ is the number of boxes of the Young diagram of $\textbf R$.

What we will however be interested in is the representations of the group $\UU(N_f)/\Z_{N}$. This puts another constraint that the charge must be a multiple of $N$ as well, i.e.
\be
q=0\bmod N\;.
\ene

Let us now couple the free fermion to the background gauge field of $F$. We denote the $\su(N_f)$-valued gauge field by $A_f = A_f^a T^a$, where $T^a$ are generators of $\su(N_f)$ in the fundamental representation, and the $\u(1)$-valued gauge field by $\mathcal{A}$. Then the Lagrangian of the system becomes
\begin{equation}
  \label{eq:UNf-free-fermion-Lagrangian}
  \mathcal{L} = i \bar{\psi} \left( \slashed{\partial} - i \slashed{A}_f^a T^a_{\textbf{R}} - i q \slashed{\mathcal{A}} \right)\psi,
\end{equation}
where $T^a_{\textbf{R}}$ are generators of $\su(N_f)$ in the representation $\textbf{R}$. As in the $\U$ case, the fermion regulators induce the Chern--Simons terms which can be written as a 4d action as
\begin{equation}
  \label{eq:UNf-CS-ind}
S_{reg.} = \frac{i s}{4\pi} \int_{X_4} \Tr \left( F_f \wedge F_f \right) + \frac{i t}{4\pi} \int_{X_4} \mathcal{F} \wedge \mathcal{F},
\end{equation}
where $F_f$ and $\mathcal{F}$ are field strengths for the gauge fields $A_f$ and $\mathcal{A}$,and $X_4$ is a spin 4-manifold whose boundary is the 3-manifold the theory lives in. In our convention, the induced Chern--Simons levels $s$ and $t$ are given by
\begin{equation}
\label{eq:induced-CS-levels-irrep}
s = -d(\textbf{R}), \qquad t = -\frac{1}{2}q^2 \text{dim} (\textbf{R})\;.
\end{equation}
If the induced Chern--Simons term can be canceled by a counterterm, the theory has no parity anomaly. This happens when the 4d action $S_{reg.}$ is independent of the choice of $X_4$ and can be written purely as a 3d action. If there were no discrete quotients in the global symmetry $F$, $A_f$ and $\mathcal{A}$ would be bona fide $\SU(N_f)$ and $\U$ gauge fields. Consequently, $S_{reg.}$ would be a boundary term whenever $s$ and $t$ are integers, and we could use $s$ and $t$ modulo 1 as indices to characterize the anomaly. However, this is no longer the case with the global symmetry $F$ that we are interested in. The presence of the discrete quotients prompts us to define a new set of indices to characterize the parity anomaly.

The new indices that can determine whether there is a parity anomaly when the global symmetry group is $\UU(N_f)/\Z_N$ are mod 2 indices $I_1$ and $I_2$, given in terms of the effective Chern--Simons levels $s$ and $t$ by
\begin{equation}
  \label{eq:indices-defn}
  \begin{split}
    I_1 &= 2\frac{N_fs-t}{N_f^2} \bmod 2,\\
    I_2 &= \frac{2 t}{N\LCM(N,N_f)} \bmod 2.
  \end{split}
\end{equation}
To see this, we write the gauge field $\mathcal{A}$ as
\begin{equation}
\mathcal{A} = \frac{\mathcal{A}_f}{N_f}+\frac{\mathcal{A}_c}{N},
\end{equation}
where $\mathcal{A}_f$ and $\mathcal{A}_c$ are properly quantized $\U$ gauge fields (with fields strengths $\mathcal{F}_f$ and $\mathcal{F}_c$, respectively) due to the $\Z_{N_f}$ and $\Z_N$ quotient\footnote{The decomposition into $\mathcal A_f$ and $\mathcal A_c$ is not unique.}. Moreover, due to the quotient $\Z_{N_f}$ between $\U$ and $\SU(N_f)$, the combination
\begin{equation}
\hat{A}_f = A_f + \frac{1}{N_f}\mathcal{A}_f\,\mathbf{I}_{N_f},
\end{equation}
whose field strength will be denoted by $\hat{F}_f$, is a properly quantized $\UU(N_f)$ gauge field. The induced Chern--Simons action then becomes
\begin{equation}
  \begin{split}
    S_{reg.} &= \frac{i t}{4\pi N^2} \int_{X_4} \mathcal{F}_c\wedge \mathcal{F}_c + \frac{i t}{2\pi NN_f}\int_{X_4} \mathcal{F}_c \wedge \mathcal{F}_f + \frac{i t}{4\pi N_f^2} \int_{X_4} \mathcal{F}_f \wedge \mathcal{F}_f\\
    & \qquad \qquad +\frac{i s}{4\pi} \int_{X_4} \Tr \left(F_f \wedge F_f\right) \\
    &= \frac{i t}{4\pi N^2} \int_{X_4} \mathcal{F}_c\wedge \mathcal{F}_c + \frac{i t}{2\pi NN_f}\int_{X_4} \mathcal{F}_c \wedge \mathcal{F}_f + \frac{i t}{4\pi N_f^2} \int_{X_4} \mathcal{F}_f \wedge \mathcal{F}_f\\
    & \qquad \qquad +\frac{i s}{4\pi} \int_{X_4} \Tr \left(\hat{F}_f \wedge \hat{F}_f\right) - \frac{i s}{4\pi N_f} \int \mathcal{F}_{f}\wedge \mathcal{F}_f  \;,
  \end{split}
\end{equation}
where we use the fact that $F_f$ is traceless to go from the first line to the second line. Now, since $\mathcal{A}_f$, $\mathcal{A}_c$ are properly quantized $\U$ gauge fields and $\hat{A}_f$ a properly quantized $\UU(N_{f})$ gauge field, the integrals
\begin{equation}
\frac{1}{4\pi} \int_X \Tr \hat{F}_f \wedge \hat{F}_f,\quad \frac{1}{4\pi}\int_X \mathcal{F}_c \wedge \mathcal{F}_c,\quad  \frac{1}{4\pi}\int_X \mathcal{F}_f \wedge \mathcal{F}_f,\quad \frac{1}{2\pi}\int_X \mathcal{F}_c \wedge \mathcal{F}_f \;,
\end{equation}
all evaluate to integral multiples of $2\pi$ on a closed spin 4-manifold $X$. Thus, $S_{reg.}$ becomes a boundary term when
\begin{equation}
s \in \Z, \quad N_fs -t \in N_f^2 \Z, \quad t \in N^2 \Z, \quad t \in NN_f \Z.
\end{equation}
The last two conditions means $t/N$ is an integral multiple of $N$ and $N_f$. These conditions can then be combined to read $t \in N \LCM(N,N_f) \Z$. Thus, there is no parity anomaly when the induced Chern--Simons levels $s$ and $t$ satisfy
\begin{equation}
s = 0 \bmod 1, \qquad N_fs-t = 0 \bmod N_f^2, \quad \text{and} \quad t = 0 \bmod N\LCM(N,N_f).
\end{equation}
However, since $s$ is always an integer whenever the last two conditions are satisfied, the condition $s\equiv 0 \bmod 1$ is not necessary. Therefore, there is no parity anomaly if and only if $N_fs-t$ is an integral multiple of $N_f^2$ and $t$ an integral multiple of $N\LCM(N,N_f)$. These conditions are equivalent to the vanishing of the indices $I_1$ and $I_2$ defined above.

The indices $I_1$ and $I_2$ of some simple representations $\textbf{R}_q$ in the cases we are interested in are summarized in Tables \ref{tab:fermionic-indices} and \ref{tab:scalars-indices}. Detailed computations for these indices are delegated to Appendix \ref{app:indices-free-fermions}. 

\subsection{The mod 16 anomaly}
\label{sec:mod-16}

Since we are interested in $\T$-preserving 3d theories, another anomaly will help us constrain the IR dynamics. The anomaly stems from the fact that Majorana fermions can seemingly be gapped only in pairs while preserving $\T$ symmetry. This is because a Majorana mass term breaks $\T$-symmetry explicitly. Consider a Majorana fermion $\xi$. Under $\T$ symmetry we can define $\xi$ to transform in two ways
\be
\T:\; \xi(x^0,x^1,x^2)\rightarrow \pm\gamma^0 (-x^0,x^1,x^2)\;.
\ene
The kinetic term $\bar\xi i\slashed\partial \xi$, where $\bar\xi=\xi^T\gamma^0$ is easily seen to be invariant under the anti-unitary $\T$ symmetry defined above. The Majorana mass term $im\bar\xi\xi$ however is not invariant under this symmetry, because of the presence of $i$, which is required to make the Minkowski action real. 

The only way to gap a free Majorana fermion is to do it in pairs, say by having $\xi_+$ transform with sign $+$ under $\T$ and $\xi_-$ with the sign $-$. Then the mass term $im\bar\xi_+\xi_-$ is invariant under the $\T$-symmetry. So, naively we may be tempted to say that the difference $n_+-n_-$ of the number of Majorana fermions $n_{\pm}$ transforming with the sign $\pm$ seems to be conserved under the RG flow, indicating that there may be a $\T$ anomaly at play. However, the arguments given above are for free fermions and so apply for weak coupling. A nonperturbative analysis \cite{Fidkowski:2013jua,Metlitski:2014xqa,Wang:2014lca,kitaev2015,Tachikawa:2016xvs} (see also \cite{Witten:2016cio}) shows that $n_+-n_-$ is conserved under the RG flow only up to $\mod $16.
In other words, we can define an index $\nu_\T$
\be
\nu_\T=n_+-n_-\bmod 16\;.
\ene
Note that the above result is only true when the $\T$-symmetry squares to $(-1)^F$ where $F$ is the fermion number, which is true for the $\T$-symmetry defined above. One can trace the origin of the number 16 from the anomaly inflow picture, where the corresponding SPT phase for this mod 16 anomaly is the 4d topological superconductor whose phases are classified by the 4th $\Pin^+$ bordism group $\Omega^{\Pin^+}_4 \cong \Z_{16}$\footnote{The appearance of the $\Pin^+$ structure is due to the fact that after Wick rotation, $T^2=(-1)^F$ becomes a reflection symmetry $R$ with $R^2=1$. This can then be used to define the $\Pin^+$ structure on a non-orientable manifold (\emph{cf.} Appendix A.2 of \cite{Witten:2015aba}).}.

Now a Dirac fermion can be thought of consisting of two Majorana fermions. We already defined a $\T$-symmetry in \eqref{eq:T_sym_dirac}. The two underlying Majorana fermions transform with the same sign under this $\T$ symmetry and hence the contribution to the $\nu_\T$ index is $\pm 2\bmod 16$.

We can also define another time-reversal symmetry\footnote{We remind the reader that in the high-energy community what we call $\T$ is usually called $CT$ and what we call $\Tp$ is usually called $\T$.} $\Tp$ by conjugating with a charge conjugation transformation
\be
\Tp:\; \psi(x^0,x^1,x^2)\rightarrow \pm\gamma^0\psi(-x^0,x^1,x^2)\;,
\ene
and define the corresponding $\nu_\Tp$ index. 
The above symmetry again squares to $(-1)^F$, and we can hence define  but this time the two underlying Majorana fermions contribute oppositely and the contribution to the corresponding $\nu_\Tp$ index is $0\bmod 16$.

\section{The $\SU(N)_0$ QCD$_{3}$(adj/f)}
\label{sec:QCDf-adj}

We will be concerned with $\T$-preserving, $\SU(N)$ gauge theories in $3d$ which always have a single Majorana fermion in the adjoint representation. If no other matter is present the theory is given by the bare action
\be\label{eq:SYM}
S=-\frac{1}{2g^2}\int \tr f^{\mu\nu}f_{\mu\nu}+ \int d^3x \bar\lambda i\slashed D\lambda+ \frac{N}{2}\frac{1}{4\pi}\int \tr\left(ada-\frac{2i}{3}a^3\right)\;,
\ene
where $a$ is the $\SU(N)$ gauge field\footnote{We will denote dynamical gauge fields with lowercase letters, and the background gauge fields with capital letters.}, $f$ is its curvature, $\lambda$ is the Majorana in the adjoint representation (viewed as an $\su(N)$-valued field), and $D_\mu=\partial_\mu-i[a_\mu,\cdot]$ is the covariant derivative in the adjoint representation. Note that we added a Chern-Simons term to cancel the the $\T$ non-invariance of the Majorana fermion measure, which transforms as
\be
\Delta S_{\lambda}= \frac{d(\textbf{adj})}{4\pi}\int \tr\left(ada-\frac{2i}{3}a^3\right)\;,
\ene
where $d(\textbf{adj})=N$ is the Dynkin index of the adjoint representation. If there are no other fermions, the bare Chern-Simons term in \eqref{eq:SYM} only makes sense if $N$ is even. The theory \eqref{eq:SYM} is a $\T$-preserving Super Yang-Mills in 3d, which was argued by Witten to break spontaneously the supersymmetry leaving a goldstino fermion behind \cite{Witten:1999ds}. Further the theory has a 1-form $\Z_N$ symmetry, which is 't Hooft anomalous, and needs to be saturated. In \cite{Acharya2001,Gomis:2017ixy}, it was proposed that in addition to a goldstino, the theory should have a decoupled TQFT $\UU(N/2)_{N/2,N}$. 

Now consider adding $N_f$ scalars $\phi^I, I=1,2,\dots, N_f$ in the fundamental representation of $\SU(N)$, which furnish a $\UU(N_f)/\mathbb Z_N$ global flavor group. We want to consider a phase where the flavor symmetry does not break at all in the Higgs phase. Before considering the case with arbitrary number of flavors, let us take one Higgs flavor and the gauge group $\SU(2)$. In that case condensing the Higgs, we can set the Higgs condensate
\begin{equation}
  \avg{\phi}= \begin{pmatrix} v\\0\end{pmatrix}\nonumber
\end{equation}
by an $\SU(2)$ gauge transformation, where $v$ is real. Note that the $\U$ flavor symmetry is not spontaneously broken because it rotates the condensate ${\phi}$ by a phase, which can be removed by an $\SU(2)$ gauge transformation.

This phase breaks the gauge symmetry completely and leaves only the $3$ massless adjoint Majorana fermions behind. However these three fermions are not gauge invariant operators and do not correspond to physical states. To construct physical states one needs to consider operators
\begin{align}
&\xi=\phi^\dagger_a {\lambda^a}_b\phi^b\\
&\psi=\phi^{a}\phi^{b}\lambda_{ab}\;,
\end{align}
where we made the color indices $a,b$ explicit and used the fact that we can raise and lower the $\SU(2)$ indices of $\lambda$ with an $\epsilon$ symbol. Note that $\xi$ is a gauge invariant real fermion operator, while $\psi$ is a complex fermion, which carries a $\U$ flavor charge. We therefore found that this phase consists of one Dirac fermion with a global $\U$ charge, which as we saw in Sec.~\ref{sec:parity-anomaly} has a mixed $\U$--$\T$ anomaly. Hence the $UV$ theory must have a parity anomaly as well. We will see how this works in Sec.~\ref{sec:anom-analysis}. Let us also comment on the real fermion $\xi$, which with a Dirac fermion makes a triplet of real fermions, furnishing an $\SO(3)$ triplet. In fact the flavor symmetry of the scalar theory has an $\SO(3)$ global symmetry, of which $\U$-baryon symmetry is a subgroup, as we discuss\footnote{This symmetry is called the custodial symmetry in the standard model electroweak sector. A quick way to see this is to turn of the $\SU(2)$ gauge fields. Then the global symmetry of scalars is $\SO(4)\cong (\SU(2)\times \SU(2))/\mathbb Z_2$. Now gauging one of the $\SU(2)$ with the dynamical gauge fields leaves $\SU(2)/\mathbb Z_2\cong\SO(3)$ global symmetry group.} in the Appendix~\ref{app:SO3_sym} (see also \cite{Choi:2018ohn}).

Now let us consider the general Higgs phase in which flavor symmetry is unbroken. This is only possible if $N_f<N$. One way to see this is to think of the Higgs field $\phi$ as an $N_f\times N$ matrix, which transforms under the color $U_c\in \SU(N)$ and the flavor $U_f\in \UU(N_f)$ as so $\phi\rightarrow U_f\phi U_c^T$. We can always choose $U_f$ and $U_c$ such that $\phi$ is diagonal\footnote{Note that because $U_c$ is special unitary, not unitary, the diagonal entries will not be real, but will have a common phase.}. Let us first consider $N=N_f$ such that $\phi$ is a square matrix, which can be taken to be diagonal. Further, in order to preserve as much flavor symmetries, we want to be able to do a simultaneous gauge and flavor rotation such that it leaves $\phi$ invariant. To allow this, let us assume that the potential for $\phi$ is such so as to be proportional to the identity matrix. Then we have
\be
\phi=v e^{i\varphi} \mathbf I\;.
\ene
Note however that the phase $\phi$ is rotated by a $\UU(N_f)$ transformation, and cannot be removed by a color rotation. Hence this phase spontaneously breaks the $\U\subset \UU(N_f)$ baryon symmetry. The order parameter is given by
\be
\phi^{I_1a_1}\dots \phi^{I_{N}a_n}\epsilon_{a_1a_2\dots a_N}\epsilon_{I_1I_2\dots I_N}\;.
\ene
Now consider $N_f<N$. In that case the $\phi$-matrix is rectangular with more columns than rows. We want to nestle an identity matrix into it to preserve as much flavor symmetries as possible, i.e.
\be
\phi = ve^{i\varphi} \begin{pmatrix}
\mathbf I_{N_f}\; \mathbf{0}_{N_f\times(N-N_f)}
\end{pmatrix}\; .
\ene
Now it is easy to see that for any flavor rotation $U_f\in \UU(N_f)$ we can choose a matrix $U_c\in \SU(N)$ as
\be
U_c= \begin{pmatrix}
U_f^*& \mathbf 0\\
\mathbf 0 &e^{i\alpha} \mathbf I
\end{pmatrix}
\ene
where $\alpha$ is chosen such that $\det U_c=1$. Then it is easy to see that $U_f \phi U_c^T=\phi$, and the Higgs condensate is invariant, implying that flavor symmetries are not broken. This is sometimes referred to as color-flavor locking. 

Now it becomes obvious that if $N_f\ge N$ we always have to have that some part of the flavor symmetry is spontaneously broken in the Higgs phase, because the gauge symmetries are insufficient to eat all the would-be Goldstone bosons. The marginal case $N_f=N-1$ where flavor symmetry need not be spontaneously broken is special because it also Higgses the whole $\SU(N)$ gauge group and is completely semiclassical. In fact we already saw one case of this when $N_f=1$ and $N=2$, and saw that the IR theory has a single free Dirac fermion. In the general case the gauge group is completely broken and there are $N^2-1$ free Majorana fermions in the IR coming from the Majorana-adjoint $\lambda$. Again these are not gauge invariant states. We construct the gauge invariant fermion operators
\begin{align}
&{\xi_{I}}^J= \phi^\dagger_I \lambda\phi^J-\frac{1}{N-1}{\delta_I}^J\phi^\dagger_K \lambda \phi^K\;,\\
&\xi=\phi^\dagger_K \lambda \phi^K\\
&\psi^I = \epsilon_{I_1I_2\dots I_{N-1}}\phi^{I_1a_1}\phi^{I_2a_2}\dots \phi^{I_{N-1}a_{N-1}}\epsilon_{a_1a_2\dots a_{N-1}a} {\lambda^a}_b\phi^{I,b}
\end{align}
Now ${\xi_{I}}^J$ is a multiplet of the adjoint representation of the flavor group (dimension $N_f^2-1= N^2-2N$), $\xi$ is just a real Majorana singlet under flavor, and $\psi^I$ (real dimension $2N-2$) is in the fundamental of $\UU(N_f)/\mathbb Z_N \cong \UU(N_f)$ for $N_f=N-1$\footnote{The statement $\UU(N_f)/\mathbb Z_N \cong \UU(N_f)$ is always true when $\GCD(N,N_f)=1$. More generally $\UU(N_f)/\Z_N\cong \UU(N_f)/\Z_{N'}$ whenever $\GCD(N,N_f)=\GCD(N',N_f)$.}. We have therefore found that the IR physics has a real adjoint fermion multiplet and a fundamental fermion multiplet of $\UU(N_f)$ flavor group, along with the a Majorana fermion singlet. Let us compute the anomaly indices $I_1$ and $I_2$ of these fermions. We have that an adjoint fermion measure contributes only to the $\SU(N_f)$ Chern-Simons level which we labeled with $s$ and gives
\be
s=\frac{N_f}{2} \bmod 1\;,
\ene 
with no contribution to the $\U$ Chern-Simons level $t$.
So the indices $I_1$ and $I_2$ are given by
\begin{align}
&I_1(\textbf{adj})=2\frac{sN_f-t}{N_f^2} \bmod 2= 1\bmod 2\;,\\
&I_2(\textbf{adj})=2\frac{t}{N\LCM(N,N_f)}\bmod 2=0\bmod 2
\end{align}
The fundamental $\psi^I$ contributes $s=1/2\bmod 1$ and $t=\frac{N_fN^2}{2}\bmod 1$\footnote{Recall that our convention is that each scalar carries a unit baryon charge, so that $\psi^I$ has baryon charge $N$.}, so we have
\begin{align}
&I_1(\textbf{f})=1\bmod 2\;\\
&I_2(\textbf{f})=1\bmod 2\;.
\end{align}
The $\xi$ fermion does not contribute to $I_{1,2}$ indices. The total indices of the Higgs theory with $N_f=N-1$ are
\begin{align}
I_1= 0\bmod 2\;\\
I_2=1\bmod 2\;.
\end{align}

What we have seen is that the theory with the Higgs has a mixed $\T$--flavor anomaly, and the theory cannot be gapped without breaking the flavor symmetry. In fact even just the $\U$ subgroup of the flavor group $\UU(N_f)$ already has a mixed anomaly with $T$, as can be seen by noticing that since $N$ has to be even to preserve the $\T$-symmetry, then $N_f$ is odd. Hence $\psi^I$ can be seen as odd number of Dirac fermions carrying a $\U$ charge. 

\subsection{Interlude: the $\mod 16$ indices}
\label{sec:interlude-mod-16}

Here we fix the $\nu_\T$ and $\nu_\Tp$ mod $16$ indices of the theory. For the purpose of this section we will take $N_f$ to be the number of fundamental fermion flavors, and keep the fundamental scalar arbitrary, as nothing depends on them. Firstly let us discuss the $\nu_{\T}$ anomaly. We have an adjoint Majorana, consisting of $N^2-1$ Majorana fermions, one for each generator of $\SU(N)$. On top of  that we also have $2N N_f$ fundamental fermions. The total $\nu_\T$-index is given by
\be\label{eq:nu_NNf}
\nu_\T = N^2-1+2N N_f \bmod 16\;.
\ene
The last term comes from taking into account $NN_f$ Dirac fermions. Note that the index is potentially gauge dependent, as we could combine the $\T$ symmetry with the gauge transformation (in the fundamental representation) of the type $U=\diag(-1,-1,\dots,-1,1,\dots 1)$ where there are $2k$ entries of $-1$ and $N-2k$ entries of $1$ (see \cite{Gomis:2017ixy} for a related discussion). Acting on the adjoint fermion $\lambda^{a}_b$, viewed as an $N\times N$ matrix, the gauge transformation leaves the sign of $(N+2k)^2+(2k)^2-1$ (the $-1$ is because the matrix ${\lambda^{a}}_b$ is traceless) entries invariant and flips $2(2k)(N-2k)$ entries. Hence it contributes to the change of the $\nu_\T$-index by $8k(N-2k) \bmod 16=8k N\bmod 16$.  Further $2k N_f$ fundamental fermions will flip sign, so that it will contribute $8 k N_f$ to the change of $\nu_\T$. So the total change of the $\nu_\T$ index is
\be
\Delta\nu_\T = 8k(N+N_f) \bmod 16 =0\bmod 16\;,
\ene
where we used the fact that $N+N_f$ is even. So the $\nu_T$-index is not ambiguous. 

Further we can rewrite the equation \eqref{eq:nu_NNf} as
\be
\nu_\T = (N+N_f)^2-1-N_f^2 \bmod 16 = 1-2(-1)^{\frac{N+N_f}{2}}-N_f^2\bmod 16\;.
\ene
The above formula can be further simplified if $N_f$ is even, then we have 
\be
\nu_\T = 3-2(-1)^{\frac{N+N_f}{2}}-2(-1)^{\frac{N_f}{2}}\bmod 16\;, \qquad N_f=0\bmod 2\;.
\ene
When $N_f$ is odd, we distinguish two cases $N_f=1\bmod 4$ and $N_f=3\bmod 4$. We have that
\be
\begin{split}
&\nu_\T = 4-4(-1)^{\frac{N_f-1}{4}}-2(-1)^{\frac{N+N_f}{2}}\bmod 16\;, \qquad N_f=1\bmod 4\;,\\
&\nu_\T = -4-4(-1)^{\frac{N_f-3}{4}}-2(-1)^{\frac{N+N_f}{2}}\bmod 16\;, \qquad N_f=3\bmod 4\;,\
\end{split}
\ene

Finally, we can also define the $\nu_\Tp$-index. In this case the Dirac fermions do not contribute at all, and only the Cartan piece\footnote{The Majorana-adjoint is algebra-valued, which we can decompose into Cartan and non-Cartan generators. Recall that $\Tp=C\T$, where $C$ is the complex-conjugation. Under $C$ the roots of the Lie group flip sign. Cartan generators are all invariant under $C$, while the non-Cartan can be combined into a combination which transforms with a $+$ and a $-$. Hence under $\Tp$ all non-Cartan components will cancel each other's contribution to the $\nu_\Tp$-index, and only Cartan components will contribute.} of the Majorana-adjoint fermion contribute to the $\nu_\Tp$ index, so that
\be
\nu_\Tp= N-1\bmod 16\;.
\ene

\subsection{The full parity anomaly analysis}
\label{sec:anom-analysis}

The same analysis for the mixed $\T$-flavor anomaly can be carried out for the general case, similar to what was discussed right before Sec.~\ref{sec:interlude-mod-16}. As alluded to at the beginning of Sec. \ref{sec:QCDf-adj}, the faithful global symmetry of QCD$_3(\textbf{adj}/\textbf{f})$  is given by
\begin{equation}
\label{eq:global-symmetry}
F = \frac{\UU(N_f)}{\Z_N} \;.
\end{equation}
Naively, the group (both gauge and global) acting on the fields in the Lagrangian is $\SU(N) \times SU(N_f) \times \U$. However, the group does not act faithfully on the fields. Through its action on the fundamental fields, we have the identifications
\begin{equation}
(u,v,\ee^{i\theta}) \sim (u \ee^{2\pi i/N}, v, \ee^{i\theta}\ee^{-2\pi i/N}), \qquad (u,v,\ee^{i \theta}) \sim (u, v \ee^{2\pi i/N_f}, \ee^{i\theta}\ee^{-2\pi i/N_f}),
\end{equation}
where $u\in \SU(N)$ and $v\in \SU(N_f)$. The faithful full symmetry group must therefore be
\begin{equation}
  \label{eq:full-sym-struct}
K = \frac{ \U\times \SU(N_f)\times \SU(N)}{\Z_N \times \Z_{N_f}} \cong \frac{\SU(N) \times \UU(N_f)}{\Z_N},
\end{equation}
which gives the global symmetry group $F=\UU(N_f)/\Z_N$ after we quotient away the gauge symmetry group $\SU(N)$. Thus, in order to use the parity anomaly to analyze such a theory, one has to generalize the basic parity anomaly described in Sec. \ref{sec:parity-anomaly} to the case when the global symmetry is $\UU(N_f)/\Z_N$. This generalization was carried out in Sec. \ref{sec:UNf-parity} by constructing the two mod 2 indices $I_1$ and $I_2$ replacing the $\U$ Chern--Simons level modulo 1 in the case of the basic parity anomaly. In this subsection, we give a fuller analysis of both the 't Hooft and the parity anomalies in the UV for the fermionic and scalar theories.

Recall that the theory can be made invariant under the time-reversal symmetry with an appropriate bare Chern--Simons term for $\SU(N)$, 
\begin{equation}
\label{eq:bare-CS-term}
S_{CS} = \frac{i r_{bare}}{4\pi}\int_{X_4} \Tr \left( f \wedge f \right),
\end{equation}
where $f$ is the field strength of the $\SU(N)$ dynamical gauge field $a$ and $r_{bare}$ is the bare Chern--Simons level. In the scalar theory, $r_{bare} = d(\textbf{adj})/2= N/2$ as we have seen earlier. For the fermionic theory, there is an additional contribution from the $N_f$ fundamental fermions, resulting in $r_{bare} = d(\textbf{adj})+N_fd(\textbf{f}) = (N+N_f)/2$. 
To study the anomalies, we turn on the background gauge fields for the global symmetry $\UU(N_f)/\Z_N$. More specifically, we couple the theory to an $\su(N_f)$-valued background gauge field $A_f$ and a $\u(1)$-valued background gauge field $\mathcal{A}$, corresponding to the $\SU(N_f)$ and $\U$ factors in the structure group $K$, respectively. Due to the discrete quotient, these gauge fields do not have the correct normalisation to qualify as bona fide $\SU(N_f)$ and $\U$ gauge fields; their normalizations, together with that of the dynamical gauge field $a$ for $\SU(N)$, are related so that the theory is put on a $K$-bundle.

In addition to simply coupling background gauge fields to the theory, we can add the Chern--Simons counterterms
\begin{equation}
\label{eq:CS-counterterm}
S_{ct} = \frac{i s_{bare}}{4\pi}\int_{X_4} \Tr\left(F_f \wedge F_f\right) + \frac{i t_{bare}}{4\pi} \int_{X_4} \mathcal{F} \wedge \mathcal{F},
\end{equation}
with integer levels $s_{bare}$ and $t_{bare}$. If we turn on the background gauge fields in a $(\SU(N_f) \times \U)$-bundle, the integrality condition for $s_{bare}$ and $t_{bare}$ suffices to make $S_{ct}$ gauge invariant. However, since the full symmetry structure of the theory is given by \eqref{eq:full-sym-struct}, the background fields can be put in a $\UU(N_f)/\Z_N$-bundle that is not a $(\SU(N_f)\times \U)$-bundle, provided that we also put the dynamical gauge field in a non-trivial $\text{PSU}(N)$-bundle\footnote{It is still an $\SU(N)$ gauge theory because we only sum over the gauge field with this \emph{fixed} bundle in the path integral.}. The bare Chern--Simons levels $r_{bare}$, $s_{bare}$, $t_{bare}$, are then subject to more stringent conditions. These conditions are \cite{Benini:2017dus} (see also Appendix \ref{app:CS-levels-derivation} for the derivation, which is similar to the derivation of the indices $I_1$ and $I_2$ in Sec. \ref{sec:UNf-parity})
\begin{equation}
\label{eq:CS-levels-conditions}
Nr_{bare}-t_{bare} \in N^2\Z, \qquad N_f s_{bare} -t_{bare} \in N_f^2 \Z, \qquad t_{bare} \in NN_f \Z,
\end{equation}
If we plug in the third requirement into the first we have that there must exist integers $a,b$ such that
\begin{equation}
 a N + bN_f = r_{bare},
\end{equation}
which hold if and only if\footnote{To see this let us assume that $r=ng+q$, where $g=\GCD(N,N_f)$, $n\in \mathbb Z$, $q\in \mathbb Z$ and $0<q<g$. Since $g=\GCD(N,N_f)$, there have to exist integers $\tilde a N+\tilde b N_f = g$, and, moreover, $g$ is the smallest number that can appear on the right hand side of this equation for any $\tilde a,\tilde b$, at least one of which is not zero. But then there would exist integers $a',b'$ such that $a' N+b' N_f=q<g$ invalidating the assumption that $r$ is not a multiple of $g$.}
\begin{equation}
  \label{eq:thooft-free-cond}
  r_{bare} = 0 \bmod \GCD(N,N_f) .
\end{equation}
If this condition is not satisfied, then it is not possible to couple the theory to the background gauge fields for the global symmetry group $F= \UU(N_f)/\Z_N$, even with the possibility of adding Chern--Simons counterterms; this is a pure flavor 't Hooft anomaly in $F$. 't Hooft anomalies of this kind have been used to establish IR dualities between different UV theories \cite{Benini:2017dus}. In this paper, however, we only consider theories with no pure flavor 't Hooft anomaly and will focus on the mixed flavor--$\T$ 't Hooft anomalies.

With the properly normalized bare Chern--Simons levels, we can now determined whether the theory remains invariant under $T$ or not by looking at the effective Chern--Simons levels
\begin{equation}
r = r_{bare}+r_{reg.}=0, \quad s = s_{bare} + s_{reg.}, \qquad t = t_{bare} + t_{reg.}.
\end{equation}
The theory is $T$-invariant when all effective Chern--Simons level vanish. However, since we can shift the counterterms $s_{bare}$ and $t_{bare}$, the levels $s$ and $t$ are not invariant quantities; we must look at $s$ and $t$ \emph{modulo} these shifts. Suppose that the shifts in the bare Chern--Simons levels are $\Delta r=0$, $\Delta t$, $\Delta s$. Again, these levels must define a properly quantized Chern--Simons term and so must satisfy \eqref{eq:CS-levels-conditions}:
\begin{equation}
\Delta t \in N^2 \Z, \qquad \Delta t \in NN_f \Z, \qquad N_f \Delta s -\Delta t \in N_f^2 \Z.
\end{equation}
In exactly the same way as in Sec. \ref{sec:UNf-parity}, the first two conditions can be combined to yield $\Delta t \in N\LCM(N,N_f)\Z$. We can see that $t$ is invariant modulo $N\LCM(N,N_f)$ and $N_fs-t$ is invariant modulo $N_f^2$. Thus, the invariant quantities that capture the parity anomaly are the indices
\begin{equation}
I_1 = 2\frac{N_fs-t}{N_f^2} \bmod 2 \quad \text{and} \quad I_2 = \frac{2t}{N\LCM(N,N_f)} \bmod 2,
\end{equation}
as defined by \eqref{eq:indices-defn}.
  
\subsubsection*{The fermionic theory}

Let us now consider more specifically the fermionic theory, where the $N_f$ fundamental fields are fermions. First, we will determine the conditions on $N$ and $N_f$ such that the theory is $T$-symmetric and the global symmetry $F$ has no 't Hooft anomaly. As we have discussed, when the theory is assumed to be $\T$-symmetric, the bare Chern--Simons level $r_{bare}$ is given by $(N+N_f)/2$. As this bare Chern--Simons level must be an integer, we either need both $N$ and $N_f$ to be even or both to be odd. To eliminate the 't Hooft anomaly, we need to satisfy \eqref{eq:thooft-free-cond}, so we need $(N+N_f)/2$ to be a multiple of $\GCD (N,N_f)$ to eliminate the 't Hooft anomaly. Together with the assumption that $N+N_f$ is even to make the theory $\T$-invariant from the get-go, one can then show that the theory is $T$-symmetric with no 't Hooft anomaly in $F$ if and only if $N$ and $N_f$ have the same number of factors of $2$.

The theory has the parity anomaly given by the non-vanishing pair of indices $I_1$ and $I_2$:
\begin{equation}
I_1 = 0 \bmod 2, \qquad I_2 = 1 \bmod 2.
\end{equation}
To see this, recall that after coupling the theory to the background gauge fields of the global symmetry, there are induced Chern--Simons levels $s_{ind}$, $t_{ind}$ for both $A_f$ and $\mathcal{A}$ due to the fundamental fermions' regulators. In our convention, they are given by
\begin{equation}
\label{eq:fermionic-bg-ind-CS}
 s_{reg.} = -\frac{N}{2}, \qquad t_{reg.} = -\frac{1}{2} NN_f \;.
\end{equation}
Next, we have to determine the levels $s_{bare}$ and $t_{bare}$ for the Chern--Simons counterterms. As we have discussed, they must satisfy
\begin{equation}
\frac{N(N+N_f)}{2} -t_{bare} \in N^2 \Z, \qquad N_f s_{bare}-t_{bare} \in N_f^2 \Z, \qquad t_{bare} \in NN_f\Z,
\end{equation}
where we again used the fact that $r_{bare} = (N+N_f)/2$ for the fermionic theory. Since  the indices $I_1$ and $I_2$ are defined in such a way that two different pairs of $s_{bare}$ and $t_{bare}$ that satisfy these conditions give the same indices, it is enough to find just any pair of $s_{bare}$ and $t_{bare}$ that works. It is easy to verify that
\begin{equation}
s_{bare} = \frac{N}{2}\left( \frac{N}{\GCD(N,N_f)}+1 \right) \quad \text{and} \quad t_{bare} = \frac{NN_f}{2}\left( \frac{N}{\GCD(N,N_f)}+1 \right)
\end{equation}
do the job. The effective Chern--Simons levels are then given by
\begin{equation}
  \begin{split}
    s &= s_{bare}+ s_{reg.} = \frac{N^2}{2\GCD(N,N_f)},\\
    t &= t_{bare}+ t_{reg.} = \frac{N^2N_f}{2\GCD(N,N_f)} = N_f s,
  \end{split}
\end{equation}
from which we obtain the indices $I_1 = 0 \bmod 2$ and $I_2 = 1 \bmod 2$ as claimed, using the definition \eqref{eq:indices-defn} and the identity $\GCD(N,N_f)\LCM(N,N_f)= NN_f$ to arrive at the final result.

\subsubsection*{The scalar theory}

The anomaly analysis for the theory with fundamental scalar fields goes almost exactly the same as the fermionic theory, with a crucial difference that the bare Chern--Simons level for the dynamical gauge field is now $r_{bare}=N/2$ instead of $(N+N_f)/2$. The condition for the vanishing 't Hooft anomaly becomes totally different as a result. For the theory to be $\T$-symmetric, $N$ must now be even only. Then $N/\GCD(N,N_f)$ must be even and $N_f/\GCD(N,N_f)$ must be odd for the pure 't Hooft anomaly to vanish.

The scalar theory has the parity anomaly captured by the two indices $I_1$ and $I_2$, which in this case are also given by
\begin{equation}
I_1 = 0 \bmod 2, \quad \text{and} \quad I_2 = 1 \bmod 2.
\end{equation}
Since the fields that are charged under the global symmetry $F$ are scalars, there is no contribution to the Chern--Simons levels of the background fields $A_f$ and $\mathcal{A}$. We only need to consider the bare Chern--Simons levels, $s_{bare}$ and $t_{bare}$,  in the derivation of the indices. Again, they must satisfy \eqref{eq:CS-levels-conditions}, which now reads
\begin{equation}
\frac{N^2}{2}-t_{bare} \in N^2 \Z, \qquad N_f s_{bare}- t_{bare} \in N_f^2 \Z, \qquad t_{bare} \in NN_f \Z,
\end{equation}
where $r_{bare}= N/2$ is used here. One choice of $s_{bare}$ and $t_{bare}$ that works is given by
\begin{equation}
s_{bare}= \frac{N^2}{2\GCD(N,N_f)}, \qquad t_{bare} = \frac{N^2N_f}{2\GCD(N,N_f)},
\end{equation}
which gives the indices $I_1= 0 \bmod 2$ and $I_2=1 \bmod 2$ through \eqref{eq:indices-defn}.

\subsection{Possible free fermion phases}
\label{sec:free-ferm-phases}

Here we speculate on the possible IR phases of the scalar and the fermion theories. In the case of extremely massive fundamental matter\footnote{It is only possible to give a mass to fundamental matter while preserving $\T$ symmetry if the number of fundamental flavors is even, and hence $N$ is even.}, the theory is essentially in the $\T$-preserving Super Yang-Mills. Super Yang-Mills was argued to break spontaneously the supersymmetry \cite{Witten:1999ds} and to have a decoupled TQFT in addition. The massive fundamental matter will then be deconfined because of the TQFT, allowing for fractional excitations, i.e. excitations in the representation of $\UU(N_f)$ not only $\UU(N_f)/\mathbb Z_N$. What we are interested in is the phase when the fundamental matter is light, or when the scalars condense. We already saw that in the particular case when $N_f=N-1$, the Higgs phase is essentially the free fermion phase. We will here consider the possibility that the IR physics is a free fermion composite phase.

Let us remind the readers that we will only consider the cases where there is a mixed flavor--$\T$ anomaly, and not the cases where there is a pure flavor anomaly. Indeed, free fermions never have a pure flavor anomaly in 3d. To satisfy this condition we require that $N/2=0\bmod \GCD(N,N_f)$ for the bosonic theory and $(N+N_f)/2 = 0\bmod \GCD(N,N_f)$ for the fermionic theory. 

Further there is no way for us to conclusively determine whether this indeed happens, as surely the anomaly can be saturated in other ways. For example one obvious way to saturate all the anomalies is to spontaneously break the $\T$-symmetry. Another possibility would be to spontaneously break the flavor symmetry\footnote{The breaking of the flavor symmetry could probably not saturate the $\nu_\T$ and the $\nu_\Tp$ indices completely, so additional assumptions would likely be needed.}. We do not discuss these possibilities here, but just mention that they are certainly viable alternatives to free fermions. In fact all of our proposals with free fermions are sensitive to $N$, and hence if we are to have a smooth large $N$ limit, we expect that free fermion phase will not be viable, and a natural candidate is a $T$-broken phase. See also comments on this in the conclusion. Also note that if the number of fermionic flavors is sufficiently high the theory can enter a conformal regime \cite{Appelquist:1988sr,Nash:1989xx,Appelquist:1989tc}.

Recall that the UV indices $I_1$ and $I_2$ are give by
\begin{align}
&I_1=0\bmod 2\;,\\
&I_2=1\bmod 2\;.
\end{align}
for both the fermionic and the bosonic theory.

Let us look at the fermionic theory first. Consider the $\nu_\T$ index of the fermionic theory we have
\be
\nu_\T= N^2-1+2NN_f\bmod 16= (N+N_f)^2+(1-N_f^2)-2\bmod 16\;.
\ene
Now notice that $(N+N_f)^2-2\bmod 16=\pm 2\bmod 16$ because $N+N_f$ must be even to preserve $\T$-symmetry. So 
\be
\nu_\T= -(N_f^2-1)\pm 2\bmod 16\;.
\ene
The above suggests that the IR physics requires a Majorana fermion in the adjoint representation contributing $\pm (N_f^2-1) \bmod 16$ and a single Dirac fermion contributing $\pm 2\bmod 16$. Indeed a Dirac fermion in the representation $\bm 1_q$ with $q=\LCM(N,N_f)$ contributes $(I_1,I_2)=(1,1)$ to the parity anomaly indices, and the adjoint Majorana fermion contributes $(I_1,I_2)=(1,0)$ giving the total index $(I_1,I_2)=(0,1)$ as it should.

The trouble is that the $\nu_\Tp$ index only sees the Majorana fermion which contributes $\nu_\Tp=N_f-1\bmod 16$, which is in general not equal to the UV index $\nu_\Tp=N-1\bmod 16$ unless $N_f=N\bmod 16$. We therefore to postulate an even number of neutral Majorana fermions to compensate for the difference, having them transform differently under $\T$ but with the same sign under $\Tp$. 

Now look at the bosonic theory. Here we have that
\be
\nu_\T= N^2-1 \bmod 16
\ene
which is either $-1$ or $3$ depending on $N$ (recall that $N$ has to be even). We can write this as $1\pm 2\bmod 16$, implying a single Majorana fermion and a Dirac fermion in the IR. Indeed the Dirac fermion has $(I_1,I_2)=(0,1)$ without the need of a Majorana adjoint in the bosonic theory. Now again we must postulate the existence of some neutral Majorana fermions in order to satisfy the anomaly $\nu_\Tp$. 

Let us now consider a scalar theory with $N_f<N-1$ in the Higgs phase. Such a Higgs phase consists of a $\SU(N-N_f)$ gauge theory coupled to $N_f$ fundamental fermion, a singlet Majorana fermion $\xi=\phi^\dagger_K\lambda\phi^K$ and an adjoint Majorana ${\Lambda_I}^J=\phi^\dagger_I \lambda\phi^J-\frac{1}{N_f}{\delta_{I}}^J\xi$. The $\SU(N-N_f)$ gauge theory coupled to $N_f$ fermions was postulated to reduce to Dirac fermion with the minimal baryon charge, and a second Majorana adjoint ${\tilde\Lambda_I}^J$ of the flavor group, as well as some extra neutral Majorana fermions. However note that the two Majorana adjoint must transform oppositely under both $\T$ and $\Tp$ in order to match the $\nu_\T$ and $\nu_\Tp$ indices, so there is nothing preventing the two from lifting each other. This would match all our assumptions above, and is further evidence that the Majorana adjoint is necessary in the IR theory of the the fermionic theory. 

While the above discussion presents a self-consistent proposal, alternative possibilities certainly exist. In fact we saw that in the fully Higgsed example with $N_f=N-1$, we have a theory with a fundamental $\UU(N_f)$ fermion, a real adjoint and a neutral adjoint. It is certainly possible to come up with other proposals for certain values of $N$ and $N_f$ which will do the job. For example, by glancing at Tables \ref{tab:fermionic-indices} and \ref{tab:scalars-indices}, we can match the parity anomaly with only a single Dirac fermion in a rank-2 (either symmetric or anti-symmetric) representation of $\SU(N_f)$ with odd baryon charge in the scalar theory with $\GCD(N,N_f) = 2$. Another example, this time in the fermionic theory when $N$ and $N_f$ are coprime, is to saturate the parity anomaly with two Dirac fermions, one in the representation $\textbf{1}_q$ with $q$ an odd multiple of $N$, another in the representation $\textbf{f}_q$ with $q$ an even multiple of $N$. Again, a number of neutral Majorana fermions must be postulated to satisfy the $\nu_T$ and $\nu_{\Tp}$ anomalies in both examples.

\begin{table}[h]
  \centering
  \begin{tabular}{|c|c|c||c|c|c|}
    \hline
    $\SU(N_f)$ & $q/N$ & $\text{gcd}(N,N_f)$ & $I_1$ & $I_2$ & $\nu_T$\\
    \hline
    $\textbf{1}$ & odd & any & $1$ & $1$ & $2 \bmod 16$\\
    $\textbf{1}$ & even & any & $0$ & $0$ & $2 \bmod 16$\\
    $\textbf{f}$ & odd & $1$ & $0$ & $0$ & $2N_f \bmod 16$\\
    $\textbf{f}$ & even &$1$ & $1$ & $0$ & $2N_f \bmod 16$\\
    ${\tiny\yng(1,1)}$ & odd & $2$ & $1$ & $1$ & $N_f(N_f-1) \bmod 16$\\
    ${\tiny\yng(1,1)}$ & even & $2$ & $0$ & $0$ & $N_f(N_f-1) \bmod 16$\\
    ${\tiny\yng(2)}$ & odd & $2$ & $1$ & $1$ & $N_f(N_f+1) \bmod 16$\\
    ${\tiny\yng(2)}$ & even & $2$ & $0$ & $0$ & $N_f(N_f+1) \bmod 16$\\
    $\textbf{adj}$ & $0$ &  any & $1$ & $0$ &$N_f^2-1 \bmod 16$\\
    \hline
  \end{tabular}
  \caption{Indices for fermions in some representations of $\UU(N_f)/\Z_N$ in the fermionic theory. The representations of $\UU(N_f)/\Z_N$ $\textbf{R}_q$ are given in terms of the $\SU(N_f)$ representations $\textbf{R}$ and the $\U$ charge $q$.}
  \label{tab:fermionic-indices}
\end{table}

\begin{table}[h]
  \centering
  \begin{tabular}{|c|c|c||c|c|c|}
    \hline
    $\SU(N_f)$ & $q/N$ & $\text{gcd}(N,N_f)$ & $I_1$ & $I_2$ & $\nu_T$\\
    \hline
    $\textbf{1}$ & odd & any & $0$ & $1$ & $2 \bmod 16$\\
    $\textbf{1}$ & even & any & $0$ & $0$ & $2 \bmod 16$\\
    $\textbf{f}$ & odd & $1$ & $1$ & $1$ & $2N_f \bmod 16$\\
    $\textbf{f}$ & even &$1$ & $1$ & $0$ & $2N_f \bmod 16$\\
    ${\tiny\yng(1,1)}$ & odd & $2$ & $0$ & $1$ & $N_f(N_f-1) \bmod 16$\\
    ${\tiny\yng(1,1)}$ & even &$2$ & $0$ & $0$ & $N_f(N_f-1) \bmod 16$\\
    ${\tiny\yng(2)}$ & odd & $2$ & $0$ & $1$ & $N_f(N_f+1) \bmod 16$\\
    ${\tiny\yng(2)}$ & even &$2$ & $0$ & $0$ & $N_f(N_f+1) \bmod 16$\\
    $\textbf{adj}$ & $0$ &  any & $1$ & $0$ &$N_f^2-1 \bmod 16$\\
    \hline
  \end{tabular}
  \caption{Indices for fermions in some representations of $\UU(N_f)/\Z_N$ in the scalar theory. The representations of $\UU(N_f)/\Z_N$ $\textbf{R}_q$ are given in terms of the $\SU(N_f)$ representations $\textbf{R}$ and the $\U$ charge $q$.}
  \label{tab:scalars-indices}
\end{table}

\section{Conclusion and outlook}
\label{sec:conclusion}

In this paper we considered 3d $\SU(N)$ Yang-Mills theories with one adjoint fermion and either bosonic or fermionic flavor matter. We showed that such systems have interesting mixed flavor--$T$ 't Hooft anomalies -- parity anomalies -- which constrain the IR physics. Such anomalies are in certain case -- when there is no pure flavor anomalies -- possible to saturate by massless composite fermions. The anomalies are characterized by two $\mod 2$ indices $I_1$ and $I_2$ described in Sec.~\ref{sec:parity-anomaly}. While the anomalies can be matched in multiple ways, we proposed some scenarios saturated by composite fermions saturating $I_{1,2}$ indices as well as the $\mod 16$ indices $\nu_{\T,\Tp}$. 

An interesting application of these anomalies is in the understanding of minimally supersymmetric $\T$-preserving Super QCD theories in 3d, which are exactly of the sort discussed here. Such theories are significantly less understood than their 4d counterparts because of the lack of holomorphy. 

Let us also comment briefly on the large $N$ limit. In this limit we do not expect the mixed flavor-$\T$ anomalies in the pure fermionic theories to be saturated by free fermions. Indeed consider the case of $\SU(N)$ with one fundamental flavor and one adjoint flavor. In this case we saw that free fermions require neutral Majorana fermions to saturate the $\nu_\Tp$ index. But this means that the free fermions are inconsistent with the smoothness of the large $N$ limit. Further, the mixed $\U_B-\T$ anomaly requires a Dirac fermion charged under $\U_B$, and such but no such state exists in cut of the the planar limit of the current-current correlator (see discussion in \cite{Eichten:1985fs}). Another option is to have spontaneous breaking of $\U_B$ symmetry, but this seems unlikely as it requires a condensation of an operator with a very large classical dimension. A natural way to saturate anomalies then is to break $\T$ symmetry spontaneously. In this case the domain walls of the $\T$ related vacua would carry anomaly inflows which are likely saturated by chiral fermions. We leave the exploration of such domain wall theories for the future.

We conclude by making some brief comments on similar anomalies in 4d. Consider the Super Yang-Mills in 4d, with an $\SU(N)$ gauge field coupled to a single Weyl fermion in the adjoin representation. As is well known by now it has a mixed 't Hooft anomaly between a discrete $\mathbb Z_N^\chi$ chiral symmetry due to the quantization of the instanton number and $\mathbb Z_N^{[1]}$ 1-form symmetry \cite{Gaiotto:2017tne}. The anomaly is saturated by spontaneous breaking of the discrete chiral symmetry. Now if we couple the model to massive fundamental matter (bosonic or fermionic), similar considerations to those of this paper will lead to the conclusion that there is a mixed anomaly between the discrete chiral symmetry and the flavor symmetry. A simple way to see that is to observe that if one activates $\U$ baryon gauge fields, one can activate the ``1-form symmetry'' fractional 't Hooft fluxes, causing the chiral symmetry to break. Note that if we give a mass to adjoint fermions the chiral symmetry is explicitly broken, but we can now have a $\theta$ term, with $\T$ symmetry at $\theta=0,\pi$. It was shown in \cite{Gaiotto:2017tne} that there may be a mixed anomaly between $\T$ and flavor symmetry, which one may think is virtually the same as the mixed $\mathbb Z_N^\chi$--flavor anomaly we discuss here. However the anomaly between $\T$ and flavor symmetry is weaker. In particular for a single fundamental fermion with $\theta=\pi$ no anomaly between $\T$ and $\U$-baryon exists (see \cite{Gaiotto:2017tne} for details) and that theory can be trivially gapped. This is not the case for a theory with a single adjoint Weyl fermion and a single fundamental fermion or boson where there is a mixed $\Z^\chi_N$--$\U$ mixed anomaly. In this respect the anomaly is akin to the anomalies discussed in \cite{Anber:2019nze,Anber:2021lzb}.

The anomaly puts constraints on the IR physics, requiring either the gapless phase, chiral symmetry breaking or flavor symmetry breaking. For massive matter, we already know that discrete chiral symmetry is broken and there are $N$ vacua. The anomaly in question, being the mixed anomaly between the chiral symmetry and flavor makes the chiral symmetry breaking robust against the deformation by light matter\footnote{If matter is scalar, then when scalars condense the theory enters a Higgs phase. Moreover if number of scalar flavors $N_s=N-1$ the gauge group can be completely Higgsed without breaking the flavor symmetry like in Sec.~\ref{sec:QCDf-adj} and it can be shown that the theory has massless fermions which reproduce the anomaly. Also note that in the case of the fermionic theory the Vafa-Witten-Weingarten theorems \cite{Vafa:1983tf,Weingarten:1983uj} forbids the breaking of the vector-like flavor symmetries as well as massless composites \cite{Aharony:1995zh} (see also \cite{Tong:2021phe}).}. An interesting application of our QCD$_3(\textbf{adj}/\textbf{f})$ is the study of domain walls between the $N$ vacua of the theory, and what anomaly descends onto the domain wall. The most interesting such domain walls are those which preserve time-reversal symmetry. The elementary domain walls connecting neighbouring vacua do not preserve $\T$ symmetry except when the gauge group is $\SU(2)$. However if the theory is supersymmetric, the domain walls are BPS and there exists a composite domain wall which preserves the time-reversal symmetry, and hence would have an interesting mixed flavor--$\T$ anomaly descending on it \cite{Delmastro2020}.

\acknowledgments{We thank Mohamed Anber, Adi Armoni, Changha Choi, Pietro Benetti Genolini, Jaume Gomis, Zohar Komargodski, David Tong, Mithat \"Unsal for helpful discussion. We are also thankful to the creators of LieART Mathematica package \cite{Feger:2019tvk} which helped with the representation analysis done in this paper. Finally we thank the organizers of the workshop ``Confronting Large N, Holography, Integrability and Stringy Models with the Real World'' at the Simons Center for Geometry and Physics and ``Continuous Advances of QCD'' in Minneapolis, where some of the content of work was presented and useful discussions arose. This work is supported by the Royal Society of London. NL is also supported by an STFC consolidated grant in ‘Particles, Strings and Cosmology’.}

\appendix

\section{The $\SU(2)$ scalar }\label{app:SO3_sym}

Consider an $\SU(2)$ scalar doublet $\phi$ coupled to the $\SU(2)$ gauge field via a covariant derivative $D_\mu=\partial_\mu-i a_\mu$. The general Lagrangian is given by
\be
(D_\mu \phi)^\dagger (D_\mu\phi)+V(\phi^\dagger\phi)\;.
\ene 
There is a manifest $\U_B$ symmetry taking $\phi\rightarrow e^{i\alpha}\phi$. However we will now show that there is a large $\SO(3)$ flavor symmetry which is not manifest.

To see this note that an $\SU(2)$ doublet can be written as 4 real scalars
\begin{equation}\label{eq:SU2_doublet}
\phi = \begin{pmatrix}
\phi_A\\
\phi_B
\end{pmatrix} =\begin{pmatrix}
\xi_1+i\xi_2\\
\xi_3+i\xi_4
\end{pmatrix} \;.
\end{equation}
Now consider a free $\SU(2)$ scalar doublet theory
\begin{equation}
\partial_\mu\phi^\dagger\partial_\mu \phi= \sum_{i=1}^4 (\partial_\mu\xi_i)^2\;.
\end{equation}
The theory above has an obvious $\SO(4)\cong (\SU(2)\times \SU(2))/\mathbb Z_2$ symmetry. We want to gauge an $\SU(2)$ subgroup of $\SO(4)$. As we will see this subgroup is a normal subgroup, so when it is gauged the remaining symmetry is $\SO(4)/\SU(2) \cong \SO(3)$.

To see that the relevant $\SU(2)$ subgroup is a normal subgroup, consider a quaternion
\begin{equation}
\xi = \xi_1 \mathbb I + i \sigma^1 \xi_4-i\sigma^2\xi_3+i\sigma^3\xi_2= \begin{pmatrix}
\phi_A &\;  - \phi_B^*\\
\phi_B &\; \phi_A^*
\end{pmatrix}= \begin{pmatrix}
\phi & \;i\sigma^2 K \phi
\end{pmatrix}\;,
\end{equation}
where in the last step we wrote the quaternion in terms of the scalar doublet \eqref{eq:SU2_doublet}, with the help of the complex-conjugation operator.

The space of quaternions $\mathbb H$ is invariant under the $(\SU(2)\times \SU(2))/\mathbb Z_2$ generated by the left and right $\SU(2)$ action of the quaternions,
\begin{equation}
\xi \rightarrow U_L\xi U_R^\dagger\;, \qquad U_{L,R}\in SU(2)_{L,R}\;.
\end{equation}
Indeed one can easily check that $U_L\xi U_R^\dagger$ is a quaternion\footnote{This follows almost immediately because $\sigma^2\xi \sigma^2=\xi^*$ and $U_{L,R}^*=\sigma^2 U_{L,R}\sigma^2$.}. Moreover, the left action on $\xi$ acts as follows
\begin{equation}
\xi \rightarrow U \xi = \begin{pmatrix}
U\phi &\; i\sigma^2 K U\phi
\end{pmatrix}\;,
\end{equation}
where note that $i\sigma^2 K U= Ui\sigma^2 K$.

Hence we identify the left action on the quaternion $\xi$ as the $\SU(2)$ symmetry that we wish to gauge. Moreover note that the left-action is quite clearly a normal subgroup, so gauging it leaves a quotient $SO(3)\cong SO(4)/SU(2)$.

\section{Quantization of the Chern--Simons levels for $\frac{\SU(N) \times \SU(N_f) \times \U}{\Z_N\times \Z_{N_f}}$}
\label{app:CS-levels-derivation}

Consider the gauge fields of a $\frac{\SU(N)\times \SU(N_f)\times \U}{\mathbb Z_{N}\times Z_{N_f}}$ gauge bundle. Let $\psi$ be the field carrying the color index of $\SU(N)$, the flavor index of  $\SU(N_f)$ index and the charge $\U$. The transformations on $\psi$ are the usual
\be
\begin{split}
&\U: \psi\rightarrow e^{i\psi}\psi\\
&\SU(N):\psi\rightarrow U\psi\;, U\in \SU(N)\\
&\SU(N_f):\psi\rightarrow U\psi\;, U\in \SU(N_f)\;.
\end{split}
\ene
The covariant derivative acting on $\psi$ is given by
\be
D_\mu=\partial_\mu-ia_\mu-iA_\mu^f-i\mathcal A\;,
\ene
with curvatures of $a,A^f$ and $\mathcal A$ being $f,F_f$ and $\mathcal F$.

We want to see when conditions \eqref{eq:CS-levels-conditions} are satisfied, that we have
\be\label{eqapp:rst_cons}
\frac{r}{4\pi}\int \tr(f\wedge f)+\frac{s}{4\pi}\int \tr(F_{f}\wedge F_f)+\frac{t}{4\pi}\int \mathcal F\wedge\mathcal F = 0\bmod 2\pi\;,
\ene
where the the integration is over a closed spin manifold.
Naively we only demand that $r,s$ and $t$ are integers, but this is not correct because of the gauge group has a quotient by $\mathbb Z_N$ and $\mathbb Z_{N_f}$. The first of these quotients can render $\frac{1}{4\pi}\int \tr f\wedge f$ quantized in units of\footnote{We consider only spin manifolds.} $\frac{2\pi}{N}$, but only if there is a corresponding fractional quantization in the $\U$ part. Similarly $\frac{1}{4\pi}\int\tr F_f\wedge F_f$ can be quantized in units of $2\pi/N_f$. Because of the twists by $\mathbb Z_N$ and $\mathbb Z_{N_f}$, $\mathcal A$ is not properly quantized and $\mathcal F$ can have $1/N$ and $1/N_f$ fractional parts.

To see when is the above satisfied, we decompose $\mathcal A = \frac{A_1}{N}+\frac{A_2}{N_F}$, where $A_1$ and $A_2$ are properly normalized gauge fields. Note that there is ambiguity how this is done, because one can always shift $A_1\rightarrow N\Lambda$ and $A_2\rightarrow N_f\Lambda$ with $\Lambda$ an arbitrary $\U$ gauge field. We have that $\hat a=a+\frac{A_1}{N}\mathbf I_{N\times N}$ and $\hat A_f=A_f+\frac{A_2}{N_f}\mathbf I_{N_f\times N_f}$ are properly normalized $U(N)$ and $U(N_f)$ gauge fields. 

We now note that \eqref{eqapp:rst_cons} can be written as
\begin{samepage}
\begin{multline}\label{eqapp:rst_cons1}
\frac{r}{4\pi}\int \tr(\hat f\wedge \hat f)+\frac{s}{4\pi}\int \tr(\hat F_f\wedge \hat F_f)+\frac{t-Nr}{4\pi N^2}\int \mathcal F_1\wedge\mathcal F_1+\frac{t-N_f s}{4\pi N_f^2}\int \mathcal F_2\wedge\mathcal F_2\\+\frac{t}{2\pi NN_f}\int \mathcal F_1\wedge\mathcal F_2 = 0\bmod 2\pi\;.
\end{multline}
\end{samepage}
The first row is $0\bmod 2\pi$ whenever $r$ and $s$ are integral, which is a prerequisite. For the other terms to vanish we must have conditions \eqref{eq:CS-levels-conditions}, repeated here for convenience
\be\tag{\ref{eq:CS-levels-conditions}}
\begin{split}
&t-Nr \in N^2\mathbb Z\;,\\
&t-N_f s \in N_f^2\mathbb Z\;,\\
&t\in N N_f\mathbb Z\;.
\end{split}
\ene
This completes the proof.

\section{Indices for free Dirac fermions in some representations of $\UU(N_f)/\Z_N$}
\label{app:indices-free-fermions}

In this Appendix, we compute the $I_1$ and $I_2$ indices that characterize the mixed $\UU(N_f)/\Z_N$--$T$ anomaly for a fermion in some simple irreducible representations of the global symmetry group $F=\UU(N_f)/\Z_N$. We consider representations that are singlet under $\SU(N_f)$, representations whose $N_f$-ality under $\SU(N_f)$ is equal to $\GCD(N,N_f)$, and the adjoint representation. The fermion is assumed to be Dirac when the representation is complex, and Majorana when it is real. The results from these computations are given in Tables \ref{tab:fermionic-indices} and \ref{tab:scalars-indices}.

First, consider the case when the fermion is a Dirac fermion in the representation $\textbf{R}_q$ of $G$, where the notation means that it is in the representation $\mathbf{R}$ of the $\SU(N)$ part and has charge $q$ under the $\U$ part of $G$. For this to be a genuine representation of $\UU(N_f)/\Z_N$, we need
\begin{equation}
  q = \ell N \quad \text{and} \quad     \abs{R}-q = 0 \bmod N_f\;.
\end{equation}
The indices $I_1$ and $I_2$, defined in \eqref{eq:indices-defn}, are given in terms of the representation as 
\begin{equation}
  \label{eq:app-indices-formuiae}
  \begin{split}
    I_1 &= \frac{2}{N_f^2}\left( N_f d(\textbf{R})-\frac{1}{2}q^2 \text{dim}(\textbf{R}) \right) \bmod 2,\\
    I_2 &= \frac{q^2\text{dim}(\textbf{R})}{N\text{lcm}(N,N_f)} \bmod 2,
    \end{split}
\end{equation}
where $d(\textbf{R})$ is the Dynkin index of the representation
$\textbf{R}$ with the normalisation $d(\textbf{f})=1/2$ for the fundamental representation $\textbf{f}$. 

Going forward, we will write $g$ for the gcd between $N$ and $N_f$.

\subsection{Singlets of $\SU(N_f)$}
\label{sec:singlet}

Consider a Dirac fermion in the representation $\textbf{1}_q$ of
$G$. The charge $q$ is given by $q=\ell N = kN_f$. This means we can
write $q=\kappa L$ where $L = \text{lcm}(N,N_f)$. Writing $L=NN_f/g$, we find that the indices $I_1$ and
$I_2$ are given by
\begin{equation}
  \begin{split}
    I_1 &= -\frac{\kappa^2 (NN_f)^2}{g^2 N_f^2} \bmod 2 = -\kappa^2n^2 \bmod 2,\\
    I_2 &= \frac{\kappa^2NN_f}{g N} \bmod 2 = \kappa^2 n_f \bmod 2,
  \end{split}
\end{equation}
where $n_f := N_f/g$ and $n:=n_f/g$. In the fermionic theory, $N$ and
$N_f$ must have the same parity in order to make the pure 't Hooft
anomaly in $G$ vanish. Consequently, both $n$ and $n_f$ are odd. As we
can choose $\kappa$ to be whatever we want, the possibilities for the
indices are
\begin{equation}
(I_1, I_2) = (1,1) \bmod 2 \quad \text{or} \quad (0,0) \bmod 2,
\end{equation}
depending on the parity of $\kappa$. 

Let's now consider the scalar theory. The pure 't Hooft anomaly
vanishing condition is $N/2 = 0 \bmod g$. So we can write
$N=2g \tilde{n}$ for some integer $\tilde{n}$. By definition,
$\text{gcd}(2g\tilde{n},gn_f) = g\text{gcd}(2\tilde{n},n_f)=g$, so we
need $n_f$ to be odd and $\text{gcd}(\tilde{n},n_f)=1$.  The charge
condition now becomes $2\ell \tilde{n}=kn_f$, so $k$ is always even,
while $\ell$ can be either even or odd following our argument above in
the fermionic theory. So, the two possibilities for the indices
are
\begin{equation}
(I_1,I_2) = (0,1) \bmod 2 \quad \text{or} \quad (0,0) \bmod 2.
\end{equation}

\subsection{Representations with $\abs{\textbf{R}}=\GCD(N,N_f)$}
\label{sec:g-box}

We now work out the indices for a Dirac fermion in the complex
representation $\textbf{R}_q$ with $\abs{\textbf{R}}=g$, for $g=1$ and $g=2$.

\subsubsection*{\underline{$g=1$}}

When $g=1$ (in particular, $N$ and $N_f$
are both odd) there is only one representation
$\textbf{R}$ of $\SU(N_f)$ that we need to consider: the
fundamental representation $\textbf{f}$.

Since $d(\textbf{f})=\frac{1}{2}$ and $\text{dim}(\textbf{f}) = N_f$, as well as
$q= \ell N = kN_f+1$, the two indices $I_1$ and $I_2$ are given by
\begin{equation}
  \begin{split}
    I_1 &= \frac{N_f-(kN_f+1)^2N_f}{N_f^2}\\
    &= \frac{1-(1+kN_f)^2}{N_f}\\
    &= -k(2+kN_f) \bmod 2\;,
  \end{split}
\end{equation}
and
\begin{equation}
  \begin{split}
    I_2 &= \frac{\ell^2N^2N_f}{N \text{lcm}(N,N_f)} \bmod 2\\
    &= \ell^2 \bmod 2\;.
  \end{split}
\end{equation}
When $\ell$ is odd, $kN_f=\ell N-1$ is even
because $N$ is odd, and vice versa. And since $N_f$ is odd as well,
this means we must take $k$ to be even. Therefore, the only possibilities for the indices $I_1$ and $I_2$ in this case are
\begin{equation}
(I_1,I_2) = (0,1) \bmod 2 \qquad \text{or} \qquad (1,0) \bmod 2.
\end{equation}

Things are different in the scalar theory. In this case, the vanishing
of the pure 't Hooft anomaly requires $N/2$ to be a multiple of
$g$. In the particular case that $g=1$, we just need $N$ to be even,
without any restriction on $N_f$ apart from being odd. Then, the
charge condition $kN_f = \ell N -1$ is always odd regardless of
$\ell$, and so is $k$ because $N_f$ is odd.

To show that there is as
solution where $\ell$ is odd, we first suppose otherwise and assume that
$\ell=2\tilde{\ell}$ is even and $k$ odd. By adding $NN_f$ to both sides of
the charge condition equation, we get
\begin{equation}
(k+N) N_f = (2\tilde{\ell}+N_f)N-1,
\end{equation}
generating a new solution $\ell^{\prime} = 2\tilde{\ell}+N_f$ odd and
$k^{\prime} = k+N$ odd, contradicting our initial assumption. So it is
always possible to choose both $\ell$ and $k$ to be both odd in the
scalar theory. The possibilities of the indices in the scalar theory are
\begin{equation}
(I_1,I_2) = (1,1) \bmod 2 \qquad \text{or} \qquad (1,0) \bmod 2.
\end{equation}

\subsubsection*{\underline{$g=2$}}

Let's look at another simple example with $g=2$. In the fermionic
theory, both $N$ and $N_f$ must be even, say $N=2^pn$ and
$N_f = 2^un_f$ where $n,n_f$ are coprime. Since
$\text{gcd}(N,N_f) =2$, $\min (p,u)$ must equal $1$. Moreover, since
we need $(N+N_f)/2$ to be a multiple of $g=2$ to have a vanishing pure
't Hooft anomaly for $G$, we need both $N/g$ and $N_f/g$ to be
odd. All these imply that $p=u=1$. Hence,
\begin{equation}
  N = 2n, \qquad N_f = 2n_f, \qquad \text{gcd}(n,n_f) = 1\;.
\end{equation}

The charge $q$ of this representation is given by $q= \ell N$ such that
\begin{equation}
  \ell N = k N_f +2\;,
\end{equation}
for some integers $\ell,k$. When $\textbf{R}$ is the rank-2 totally
antisymmetric representation, we have
$\text{dim}(\textbf{R})=\frac{1}{2}N_f(N_f-1)$ and
$d(\textbf{R})= \frac{1}{2}(N_f-2)$. Then, it can be shown that
\begin{equation}
  \begin{split}
    I_1 &= 1-\frac{1}{2}(4+k N_f)(N_f-1)k \bmod 2\;,\\
    I_2&= \ell^2(N_f-1) \bmod 2\;.
  \end{split}  
\end{equation}
If we want to match the mixed T-anomaly with this representation
instead of the Dirac singlet (in addition to the adjoint Majorana), we
need to get $I_1 = I_2 = 1 \bmod 2$. This is only possible when $\ell$ is
odd and $k$ is even. But from the condition $2\ell n = 2(kn_f+1)$, we see
that if $\ell$ is odd, $k$ must be even, and vice versa. Thus, we only
need to show that there exists a solution to the condition
$\ell N=k N_f +2$ such that $\ell$ is odd.

We proceed exactly as in the $g=1$ case. Suppose to the contrary, that
a solution $(\ell,k)$ to our charge condition only exist when
$\ell=2\tilde{\ell}$ is even and $k$ odd. Then, the charge condition becomes
\begin{equation}
2\tilde{\ell}n = kn_f+1.
\end{equation}
Adding $nn_f$ to both sides of the equation, we get 
\begin{equation}
(2\tilde{\ell}+n_f)n=(k+n)n_f +1.
\end{equation}
Thus, $(2\tilde{\ell}+n_f, k+n)$ is also a solution to the charge
condition if $(\ell,k)$ is. But $2\tilde{\ell}+n_f$ is odd because
$n_f$ is odd, contradicting our initial assumption. Therefore, it is
always possible to choose $\ell$ such that it is odd. The two
possibilities of the indices are
\begin{equation}
(I_1,I_2) = (1,1) \bmod 2 \qquad \text{or} \qquad (0,0) \bmod 2\;.
\end{equation}

For the scalar theory, the vanishing 't Hooft anomaly requires
$N/2 = 0 \bmod 2$ so $N=4\tilde{n}$ for some integer
$\tilde{n}$. Moreover, since $\text{gcd}(N,N_f)$ is $2$ and not more,
we must have $N_f = 2n_f$ with $n_f$ odd and
$\text{gcd}(\tilde{n},n_f)=1$.  Then, the charge condition becomes
$kn_f=2\ell\tilde{n}-1$, telling us that $k$ is always odd in this
case. Using the same method as in all our discussion so far, we can
show that $\ell$ can be chosen to be either odd or even. Thus, the
possibilities for the indices are
\begin{equation}
(I_1,I_2) = (0,1) \bmod 2 \quad \text{or} \quad (0,0) \bmod 2\;.
\end{equation}

When $\textbf{R} = {\tiny \yng(2)}$, we obtain the same results as in
the antisymmetric case. More precisely, we use
\begin{equation}
\text{dim}({\tiny \yng(2)}) = \frac{1}{2}N_f(N_f+1) \qquad \text{and} \qquad d({\tiny \yng(2)}) = \frac{1}{2}(N_f+2)
\end{equation}
to obtain
\begin{equation}
  \begin{split}
    I_1 &= -1 -k^2 n_f(2n_f+1) \bmod 2\;,\\
    I_2 &= \ell^2 (2n_f+1) \bmod 2\;.
  \end{split}
\end{equation}
Then, from the fact that $n_f$ is odd in both the fermionic and the
bosonic cases, the two indices solely rely on the parities of $\ell$ and
$k$, which we have just analyzed in the couple preceding
paragraphs. Thus,
\begin{equation}
(I_1,I_2) = (1,1) \bmod 2 \quad\text{or} \quad (0,0) \bmod 2
\end{equation}
in the fermionic case, and
\begin{equation}
(I_1,I_2) = (0,1) \bmod 2 \quad \text{or} \quad (0,0) \bmod 2  
\end{equation}
in the scalar theory.

\subsection{The adjoint representation}

Finally, let's look at the indices from a Majorana fermion in the adjoint representation $\textbf{adj}_0$. Since it is neutral under $\U$ and is a Majorana fermion, the indices are instead given by
\begin{equation}
I_1 =  \frac{d(\textbf{adj})}{N_f} \bmod 2, \qquad I_2 = 0 \bmod 2,
\end{equation}
where the factor of 2 difference in $I_1$ compared to \eqref{eq:app-indices-formuiae} is because the Chern--Simons level induced from the regulator of a Majorana fermion is half of the Dirac fermion's contribution. The Dynkin index $d(\textbf{adj})$ of the adjoint representation of $\SU(N_f)$ in our normalization is $N_f$, so the indices contributed by an adjoint Majorana fermion are
\begin{equation}
(I_1, I_2) = (1, 0) \bmod 2,
\end{equation}
in both the fermionic and the scalar theories.

\bibliographystyle{JHEP}
\bibliography{references.bib}
\end{document}